\documentclass[final,5p,times,twocolumn]{elsarticle}

\usepackage{amssymb}
\usepackage{color}
\usepackage{amsmath}
\usepackage{multirow}
\usepackage{booktabs} 
\usepackage{bm}
\usepackage{ulem}
\usepackage{subfigure} 
\usepackage[pdfstartview=FitH, colorlinks,
linkcolor={red!60!black!},
anchorcolor={green!80!black!},
urlcolor={blue!80!black!},
citecolor={blue!50!black!}]{hyperref}
\usepackage[table]{xcolor}

\journal{Physics Letters B}

\begin{document}


\begin{frontmatter}



\title{How do mirror charge radii constrain density dependence of the symmetry energy?}


\author[ad1,ad2,ad3]{Bai-Shan Hu\corref{cor1}}

\address[ad1]{TRIUMF, Vancouver, BC V6T 2A3, Canada}
\address[ad2]{National Center for Computational Sciences, Oak Ridge National Laboratory, Oak Ridge, Tennessee 37831, USA}
\address[ad3]{Physics Division, Oak Ridge National Laboratory, Oak Ridge, Tennessee 37831, USA}
\cortext[cor1]{baishanhu4phys@gmail.com}

\begin{abstract}

It has recently been suggested that differences in the charge radii of mirror nuclei ($\Delta R^{\rm mirr}_{\rm ch}$) are strongly correlated with the neutron-skin thickness ($R_{\rm skin}$) of neutron-rich nuclei and with the slope of the symmetry energy ($L$). To test this assumption, we present ab initio calculations of $R_{\rm skin}$ in $^{48}$Ca and $^{208}$Pb, $\Delta R^{\rm mirr}_{\rm ch}$ in $^{36}$Ca$-^{36}$S, $^{38}$Ca$-^{38}$Ar, $^{41}$Sc$-^{41}$Ca, $^{48}$Ni$-^{48}$Ca, $^{52}$Ni$-^{52}$Cr, and $^{54}$Ni$-^{54}$Fe mirror pairs, and $L$. Employing the recently developed 34 chiral interaction samples, identified by the history matching approach, we conduct rigorous statistical analysis of correlations among $\Delta R^{\rm mirr}_{\rm ch}$, $R_{\rm skin}$ and $L$, accounting for quantified uncertainties from low-energy constants of chiral interaction, chiral effective field theory truncation and many-body method approximation.
The ab initio results reveal an appreciable $\Delta R^{\rm mirr}_{\rm ch}-L$ correlation in $fp$-shell mirror pairs. 
However, contrary to previous studies, the present calculation finds that the studied $sd$-shell mirror pairs do not exhibit any $\Delta R^{\rm mirr}_{\rm ch}-L$ correlation.

\end{abstract}


\end{frontmatter}



\section{Introduction}
Symmetry energy encodes the energy cost per nucleon associated with converting all protons into neutrons in symmetric nuclear matter. It characterizes the isospin-dependent part of the equation of state (EoS) for asymmetric nuclear matter. The density dependence of the symmetry energy plays a critical role in various areas, including astrophysics, nuclear structure, and nuclear reactions. For instance, it influences the structure of neutron stars, the mechanism of core-collapse supernovae, the properties of stellar nucleosynthesis, and the dynamics of heavy-ion collisions; see, e.g., Refs.~\cite{Danielewicz1592,Lattimer536,Hebeler_2013,Li2014,RevModPhys.89.015007,PhysRevLett.121.062701}. At the saturation density, the slope of symmetry energy, denoted as $L$, is one critical EoS parameter for extrapolating the symmetry energy to both lower and higher densities. However, $L$ cannot be directly measured in the laboratory.

Isovector indicators, such as neutron skin thickness, parity-violating asymmetry, and electric dipole polarizability, have been used to set boundaries on slope $L$ in the various nuclear models. Here the neutron skin thickness $R_{\rm skin}$ is defined as the difference between the root-mean-squared point radii of neutrons and protons in a nucleus. Recent PREX-II \cite{PhysRevLett.126.172502} and CREX \cite{PhysRevLett.129.042501} experiments measured the parity-violating asymmetry in the polarized elastic electron scattering at one momentum transfer and extract the $R_{\rm skin}$ in $^{208}$Pb and $^{48}$Ca using different models. $R_{\rm skin}$ in different nuclei are also extracted from proton elastic scattering \cite{PhysRevC.82.044611}, antiprotonic atoms \cite{PhysRevLett.87.082501}, and coherent pion photoproduction \cite{PhysRevLett.112.242502} measurements. The electric dipole polarizability offers an electromagnetic probe to constrain the slope $L$, but its model dependence is challenging to quantify \cite{PhysRevC.85.041302,PhysRevLett.129.232501,PhysRevC.92.064304}. In addition to terrestrial experiments, the advent of multi-messenger astronomy has opened up the possibility of exploring the nature of neutron stars and the EoS of dense nuclear matter by gravitational-wave and electromagnetic observations \cite{PhysRevLett.119.161101,Abbott_2017,PhysRevLett.120.172702,PhysRevLett.126.061101,PhysRevLett.127.192701}. Despite substantial advancements in these electroweak, hadronic, electromagnetic and gravitational-wave probes for constraining the EoS of dense nuclear matter, the value of $L$ remains considerable uncertainty, potentially ranging between 20 MeV and 143 MeV \cite{Newton2014,PhysRevC.92.064304,PhysRevLett.125.202702,PhysRevLett.126.172503,PhysRevLett.127.192701}.

Recently, it has been suggested~\cite{PhysRevLett.119.122502,PhysRevC.97.014314,PhysRevResearch.2.022035,Boso_2020,PhysRevLett.127.182503,PhysRevC.108.015802} that a difference in the charge radii of mirror nuclei, $\Delta R^{\rm mirr}_{\rm ch} \equiv R_{\rm ch}(^{A}_{Z}X_{N})-R_{\rm ch}(^A_NY_{Z})$, is proportional to both the $R_{\rm skin}$ of neutron-rich nuclei and the slope $L$ of the symmetry energy. This finding implies that $\Delta R^{\rm mirr}_{\rm ch}$ could serve as a purely electromagnetic alternative to parity-violating asymmetry and the electric dipole polarizability for constraining $L$.
Precise measurements of charge radii in $^{36}$Ca, $^{36}$S, $^{38}$Ca, $^{38}$Ar, $^{54}$Ni and $^{54}$Fe have been conducted \cite{Rch_2004,Miller2019,PhysRevLett.127.182503}. A robust correlation between $\Delta R^{\rm mirr}_{\rm ch}$ and $L$ would enable stronger constraints on $L$ compared to $R_{\rm skin}$ or electric dipole polarizability. However, recent studies~\cite{PhysRevC.105.L021301,PhysRevC.107.034319} claim that the precise data on mirror charge radii cannot provide a stringent constraint on $L$ due to large theoretical uncertainties on $\Delta R^{\rm mirr}_{\rm ch}$, such as pairing correlations. In this letter, we probe the correlations among $\Delta R^{\rm mirr}_{\rm ch}$, $R_{\rm skin}$ and $L$ from first principles and address the question, ``How robust is the mirror charge radii constraint on density dependence of the symmetry energy?''

Significant progress has been made in ab initio calculations, which start from an initial Hamiltonian with underlying nucleon-nucleon (NN) and three-nucleon (3N) interactions derived from chiral effective field theory ($\chi$EFT). These calculations can now describe light to medium-mass nuclei and even heavy nuclei like $^{208}$Pb \cite{annurev-nucl-101917-021120,10.3389/fphy.2020.00379,Hu22208Pb,PhysRevC.107.024310}.
Multi-reference in-medium similarity renormalization group (MR-IMSRG) has studied the correlation between $L$ and $\Delta R^{\rm mirr}_{\rm ch}$ using an EM family of chiral interactions, specifically EM2.0/2.0(PWA), EM2.2/2.0, EM2.0/2.0 and EM1.8/2.0 \cite{PhysRevResearch.2.022035}. This investigation resulted in a narrow range of values for both $L$ and $\Delta R^{\rm mirr}_{\rm ch}$. All of the EM family interactions are based on the Entem and Machleidt's N$^3$LO NN force with fixed parametrizations of low-energy constants (LECs). As such, one does not see clear correlations between $\Delta R ^{\rm mirr}_{\rm ch}$ and $L$ in Ref.~\cite{PhysRevResearch.2.022035}. Recently, Ref.~\cite{PhysRevLett.130.032501} has probed ab initio $\Delta R_{\rm ch}^{\rm mirr}$ for nuclei with $6 \le A \le 56$ using four commonly used chiral NN+3N interactions. However, this study does not investigate the correlation between $\Delta R_{\rm ch}^{\rm mirr}$ and $L$.

A rigorous statistical analysis of correlations between finite nuclei observables and infinite nuclear matter quantities requires the incorporation of all relevant sources of uncertainty, including the LECs of chiral interaction. In a recent work \cite{Hu22208Pb}, the history matching approach was employed to explore and reduce the vast number of $10^9$ different parameterizations within the 17-dimensional parameter space of LECs at chiral NNLO with explicit delta isobars ($\Delta$). By confronting with data in nucleon-nucleon scattering and select light nuclei, while accounting for relevant uncertainties, the study \cite{Hu22208Pb} identified the so-called non-implausible LECs domain that could not be ruled out, arriving at a probability distribution for an ensemble of interactions that accurately reproduces the bulk properties of $^{208}$Pb. These non-implausible interaction samples allow us to carry out a comprehensive examination of correlations between nuclear matter properties and observables in finite nuclei. In this letter, we utilize the 34 non-implausible chiral interaction samples created in the aforementioned study~\cite{Hu22208Pb} to probe the potential correlation among $\Delta R^{\rm mir}_{\rm ch}$, $R_{\rm skin}$ and $L$.

\section{Methods \label{sec:1}}
We compute $\Delta R^{\rm mirr}_{\rm ch}$ and $R_{\rm skin}$ for a range of commonly studied nuclei, including $^{36}$Ca, $^{36}$S, $^{38}$Ca, $^{38}$Ar,$^{41}$Sc, $^{41}$Ca, $^{48}$Ni, $^{48}$Ca, $^{52}$Ni, $^{52}$Cr, $^{54}$Ni, $^{54}$Fe and $^{208}$Pb, using the valence-space in-medium similarity renormalization group (VS-IMSRG) approach~\cite{HERGERT2016165,PhysRevLett.118.032502,Stro19ARNPS} with the 34 non-implausible chiral NN+3N interaction samples. 
Work in a 15 major-shell harmonic oscillator (HO) space, we impose an additional $E_{\rm 3max}$=28 truncation for storing 3N matrix elements \cite{PhysRevC.105.014302,Hu22208Pb}. This model space is the same as that used in the previous calculation of $R_{\rm skin}$ in $^{208}$Pb \cite{Hu22208Pb}, and we find that it provides converged results for all the nuclei studied in this work. First, we transform the Hamiltonian and squared intrinsic point-nucleon radius operators to the Hartree-Fock (HF) basis. Then, we use the VS-IMSRG to derive an approximate unitary transformation to decouple the $s_{1/2}d_{3/2}f_{7/2}p_{3/2}$ multishell valence space Hamiltonian above a $^{28}$Si core for $A$=38 and $A$=41 pairs, and the $fp$-shell valence space for $A$=52 and 54 pairs. For $^{36}$Ca, $^{36}$S, $^{48}$Ni, $^{48}$Ca and $^{208}$Pb, we perform closed-shell calculations and do not decouple a neutron (proton) valence space, as explained in Ref.~\cite{PhysRevLett.126.022501}. Note that the results for $R_{\rm skin}$ of $^{48}$Ca and $^{208}$Pb have been previously published in Ref.~\cite{Hu22208Pb}.
Applying the same unitary transformation to the squared intrinsic point-nucleon radius operator, we further construct a two-body-level effective valence-space operator consistent with the Hamiltonian. The charge radius is obtained from the point-proton radius using the standard expression \cite{PhysRevC.94.014303,PhysRevLett.128.022502}. The effects of 3N correlations between valence nucleons are captured via the ensemble normal ordering procedure \cite{PhysRevLett.118.032502}.
The final exact diagonalization of valence-space Hamiltonian is performed using the {\tt KShell} shell-model code \cite{SHIMIZU2019372}.
The ab initio results of the slope $L$ within the same 34 chiral NN+3N interaction samples are taken from Ref.~\cite{Hu22208Pb}.

The results for observable $y$, which include $\Delta R^{\rm mirr}_{\rm ch}$, $R_{\rm skin}$ and $L$, calculated using the 34 non-implausible chiral interaction samples, do not provide a probabilistic interpretation due to the absence of actual probability distributions.
To make quantitative predictions with a statistical interpretation, we extract representative samples from the posterior probability density function $\mathcal{P}(\vec{\theta} | \mathcal{D})$ using Bayes' theorem,
\begin{equation}
\mathcal{P}(\vec{\theta} | \mathcal{D}) \propto \mathcal{P}(\mathcal{D} | \vec{\theta})\mathcal{P}(\vec{\theta}),
\end{equation}
where $\mathcal{P}(\mathcal{D} | \vec{\theta})$ represents the likelihood, $\mathcal{P}(\vec{\theta})$ is the prior distribution, $\mathcal{D}$ is calibration data set, and the parameter vector $\vec{\theta}$ corresponds to the 17 LECs used in this study. We assume a uniform prior probability $\mathcal{P}(\vec{\theta})$ for all LECs $\vec{\theta}$ except $c_1-c_4$, for which the prior is adopted as a multivariate Gaussian originating from the Roy-Steiner analysis of pion-nucleon scattering data \cite{siemens2017}. The history matching procedure yields 34 non-implausible samples from this prior, and its implausibility constraints exclude samples expected to make negligible contributions to the $\mathcal{P}(\vec{\theta} | \mathcal{D})$, even though the data likelihood is not involved in this process.
We then employ the sampling/importance resampling method \cite{Smith1992,Jiang2022} of Bayes' theorem to resample a set $\{\vec{\theta}_i\}_{i=1}^{N}$ from the 34 discrete distribution  {$\{\vec{\theta^*_i}\}_{i=1}^{34}$} obtained from the history matching according to their importance weights $q_i$. This resampled set will then be approximately distributed according to $\mathcal{P}(\vec{\theta} | \mathcal{D})$. 

The importance weights are obtained by
\begin{equation}
q_i = \frac{\mathcal{P}(\mathcal{D} | \theta^*_i)}{\sum_{j=1}^{34}\mathcal{P}(\mathcal{D} | \theta^*_j)}.
\end{equation}
Here, the calibration data set $\mathcal{D}$ is defined by the binding energies and charge radii of $^{48}$Ca, as well as its structure-sensitive $2^+$ excitation energy. Note that several observables ($\mathcal{O}_{A=2,3,4,16}$) have been used in the history matching procedure: nucleon–nucleon scattering phase shifts up to an energy of 200 MeV; the binding energy, charge radius, and quadrupole moment of $^2$H; and the binding energies and charge radii of $^3$H, $^4$He, and $^{16}$O. Consequently, the 34 non-implausible interaction parameterizations, $\vec{\theta^*}$, are already confronted with these experimental data ($\mathcal{O}_{A=2,3,4,16}$), and we do not include these $\mathcal{O}_{A=2,3,4,16}$ into $\mathcal{D}$ during the sampling/importance resampling step. A normally distributed likelihood $\mathcal{P}(\mathcal{D} | \vec{\theta^*})$ is used. Ref.~\cite{Hu22208Pb} has demonstrated that the final results are not sensitive (about 1\%) to the likelihood $\mathcal{P}(\mathcal{D} | \vec{\theta^*})$ definition, as confirmed by comparisons using different definitions with a non-diagonal covariance matrix or a Student $t$ distribution with heavier tails. In this work, we find the correlation between $\Delta R^{\rm mirr}_{\rm ch}$ and $L$ (or $R_{\rm skin}$ and $L$) remains largely unchanged, even when importance weights $q_i$ are not considered.

Following this procedure, the posterior predictive distribution (PPD) dependent on the LECs ($\vec{\theta}$) is given by
\begin{equation}
\label{PPD}
\mathrm{PPD}_{\bm{\theta}}=\left\{y_k(\vec{\theta}): \vec{\theta} \sim \mathcal{P}(\vec{\theta} | \mathcal{D})\right\},
\end{equation}
where $y_k$ denotes the prediction of $y$ at truncation order $k$ in the chiral EFT expansion. In this work, $y_k$ ($\Delta R^{\rm mirr}_{\rm ch}$, $R_{\rm skin}$ and $L$) is calculated by VS-IMSRG with chiral NN+3N interaction at NNLO, corresponding to $k=3$. Finally, the full PPD is defined, in analogy with Eq.~(\ref{PPD}), as $y$, which is the sum 
\begin{equation}
\label{errors}
y=y_k+ \epsilon_{\chi\text{EFT}} + \epsilon_{\text{method}},
\end{equation}
with $\epsilon_{\chi\text{EFT}}$ and $\epsilon_{\text{method}}$ representing the $\chi$EFT truncation error and many-body method approximation error, respectively. We assume $\chi$EFT and method errors to be independent of the parameters. 
In practice, we generate $10^5$ samples for the $\Delta R^{\rm mirr}_{\rm ch}$ and $L$ (or $R_{\rm skin}$ and $L$) pairs from the bivariate PPD by resampling the 34 samples of the PPD$_{\bm\theta}$ in Eq.~(\ref{PPD}) with probability mass $q_i$ on $\vec{\theta^*_i}$, and adding samples from the error terms in Eq.~(\ref{errors}). We assume that the correlation coefficient between the errors, $\epsilon_{\chi\text{EFT}}$ and $\epsilon_{\text{method}}$, in $\Delta R^{\rm mirr}_{\rm ch}$ and $L$, as well as between $R_{\rm skin}$ and $L$, is identical to the correlation coefficient of the observables computed from the 34 non-implausible samples.

The $\chi$EFT truncation errors for observable $y$, which include $\Delta R^{\rm mirr}_{\rm ch}$, $R_{\rm skin}$ and $L$, are quantified by adopting the EFT convergence model \cite{PhysRevC.92.024005,PhysRevC.100.044001}. The order-$k$ $\chi$EFT prediction is
\begin{equation}
y_k = y_\mathrm{ref} \left( \sum_{i=0}^k c_i Q^i \right),
\label{yk}
\end{equation}
and its $\chi$EFT truncation error is
\begin{equation}
\label{eq:EFTconvergence}
\delta y_k = y_\mathrm{ref} \left( \sum_{i=k+1}^\infty c_i Q^i \right),
\end{equation} 
where observable coefficients $c_i$ are expected to be naturally sized, and the expansion parameter $Q$ is set to 0.42, following the Bayesian error model for nuclear matter reported in Ref.~\cite{Hu22208Pb}. 

We use the pointwise Bayesian statistical model \cite{PhysRevC.92.024005} from the package {\tt gsum} \cite{PhysRevC.100.044001} to quantify $\chi$EFT truncation error for $\Delta R_{\rm ch}^{\rm mirr}$.
The magnitudes of these terms are quantified by learning about the distribution for $c_i$, which is assumed to be described by a single
normal distribution per observable type with zero mean and a variance parameter $\bar{c}^2$. Specifically, we employ next-to-leading order (NLO) and next-to-next-to-leading order (NNLO) interactions from Ref.~\cite{PhysRevC.102.054301}, and compute $\Delta R_{\rm ch}^{\rm mirr}$ at both orders ($k$=2 and $3$) for all studied mirror pairs. The reference values
$y_\mathrm{ref}$ are set to $1.574 \cdot I$, based on a global investigation of $\Delta R_{\rm ch}^{\rm mirr}$ across nuclei with mass numbers ranging from 22 to 54 as detailed in Ref.~\cite{PhysRevLett.130.032501}. Here, the isospin asymmetry $I$ is defined as $(N-Z)/A$. From the NLO ($k=2$) and NNLO ($k=3$) results of $\Delta R_{\rm ch}^{\rm mirr}$, we extract
$\bar{c}^2=2.24$ and perform the geometric sum of the second term in
Eq.~\eqref{eq:EFTconvergence}. We checked that the uncertainty in $\Delta R_{\rm ch}^{\rm mirr}$ is not sensitive to the $Q$ value. 

\begin{figure}[h]
\centering
\setlength{\abovecaptionskip}{0pt}
\setlength{\belowcaptionskip}{0pt}
\includegraphics[scale=0.42]{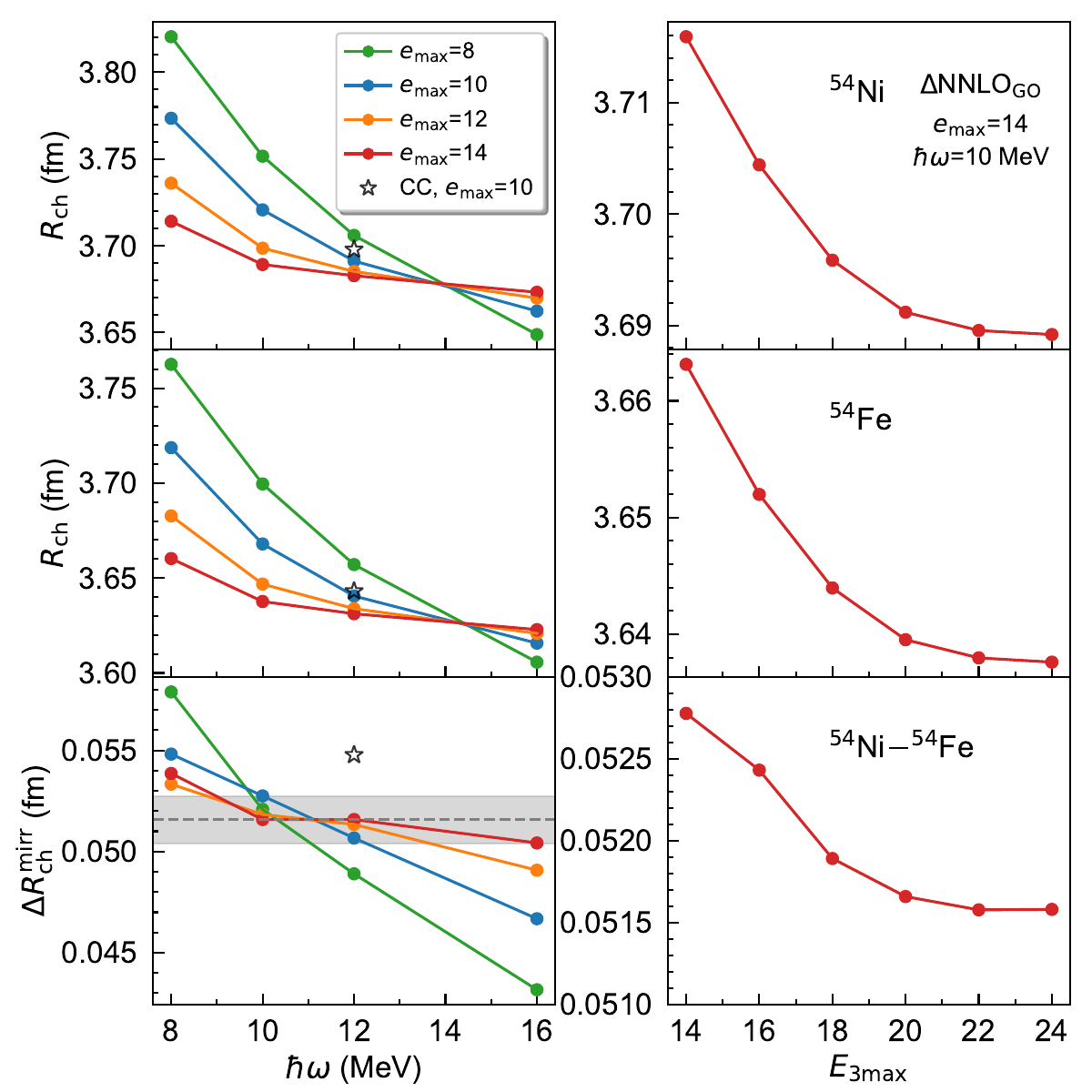}
\caption{\label{Convergence} Convergence of the charge radius ($R_{\rm ch}$) in $^{54}$Ni and $^{54}$Fe as a function of the $e_{\rm max}$ and $E_{\rm 3max}$ truncations. The star symbol denotes the result calculated using the deformed coupled cluster (CC) method \cite{Gaute,PhysRevC.102.051303,PhysRevC.105.L021303,PhysRevC.105.064311}, and the band represents the uncertainty assigned to the model space convergence.}
\end{figure}

The $\chi$EFT uncertainty of slope $L$ is quantified using the Gaussian process model, as published in Ref.~\cite{Hu22208Pb}. The $\epsilon_{\chi\text{EFT}}$ for $R_{\rm skin}$ is also taken from Ref.~\cite{Hu22208Pb}.
The many-body method approximation error $\epsilon_{\text{method}}$ is analyzed by examining model space convergence patterns and comparing the differences between IMSRG, and coupled-cluster (CC) methods. 
In particular, we compare the $\Delta R_{\rm ch}^{\rm mirr}$ results from VS-IMSRG and deformed CC calculations \cite{Gaute,PhysRevC.102.051303,PhysRevC.105.L021303,PhysRevC.105.064311} for full open-shell nuclei, as shown in Fig.~\ref{Convergence}. These $\epsilon_{\chi\text{EFT}}$ and $\epsilon_{\text{method}}$ errors are enumerated in Table~\ref{table_error}.

\renewcommand{\arraystretch}{1.5} 
\begin{table}
\centering
\caption{
\label{table_error} Error assignments for the differences in charge radii of mirror nuclei ($\Delta R^{\rm mirr}_{\rm ch}$), represented by the estimated standard deviation $\sigma$.
The final predictions from the posterior predictive distribution (PPD) shown in Fig.~\ref{Rskin_Rch} are summarized by the medians and the marginal 68\% credibility regions in the last column. Experimental values for charge radius $R_{\rm ch}$ are taken from Refs~\cite{Rch_2004,Miller2019,PhysRevLett.127.182503}.}
\begin{tabular}{ccccc p{cm}}
\hline
\hline
$\Delta R^{\rm mirr}_{\rm ch}$ (fm) & Expt  & $\sigma_{\chi \rm {EFT}}$ & $\sigma_{\rm method}$ &  PPD \\
\hline
$^{36}$Ca$-^{36}$S   &0.150$\pm$0.004  &0.008 &0.013  &$0.161^{+0.018}_{-0.018}$\\
$^{38}$Ca$-^{38}$Ar  &0.063$\pm$0.003 &0.004 &0.006 &$0.068^{+0.009}_{-0.009}$\\
$^{41}$Sc$-^{41}$Ca  &$-$ &0.002 &0.004 &$0.043^{+0.007}_{-0.007}$\\
$^{48}$Ni$-^{48}$Ca  &$-$ &0.012 &0.021 & $0.266^{+0.031}_{-0.032}$\\
$^{52}$Ni$-^{52}$Cr  &$-$ &0.007 &0.009 &$0.112^{+0.015}_{-0.015}$\\
$^{54}$Ni$-^{54}$Fe  &0.049$\pm$0.004 &0.004 &0.004 &$0.053^{+0.010}_{-0.010}$\\
\hline
\hline
\end{tabular}
\end{table}


In the present work, we assess the correlation between quantities $a$ and $b$ using the coefficient of determination (CoD) \cite{PhysRevC.105.L021301},
\begin{equation}
\label{CoD}
\text{CoD} = R^2=\left[\dfrac{\text{cov}(a,b)}{\sigma_a\sigma_b}\right]^2,
\end{equation}
where $R$ is the Pearson correlation coefficient, cov($a$,$b$) represents the covariance, $\sigma$ denotes the standard deviation.
The CoD provides information on how well the quantity $a$ is determined by $b$. 

\section{Results and discussion}

\begin{figure*}
\centering
\setlength{\abovecaptionskip}{0pt}
\setlength{\belowcaptionskip}{0pt}
\includegraphics[scale=0.68]{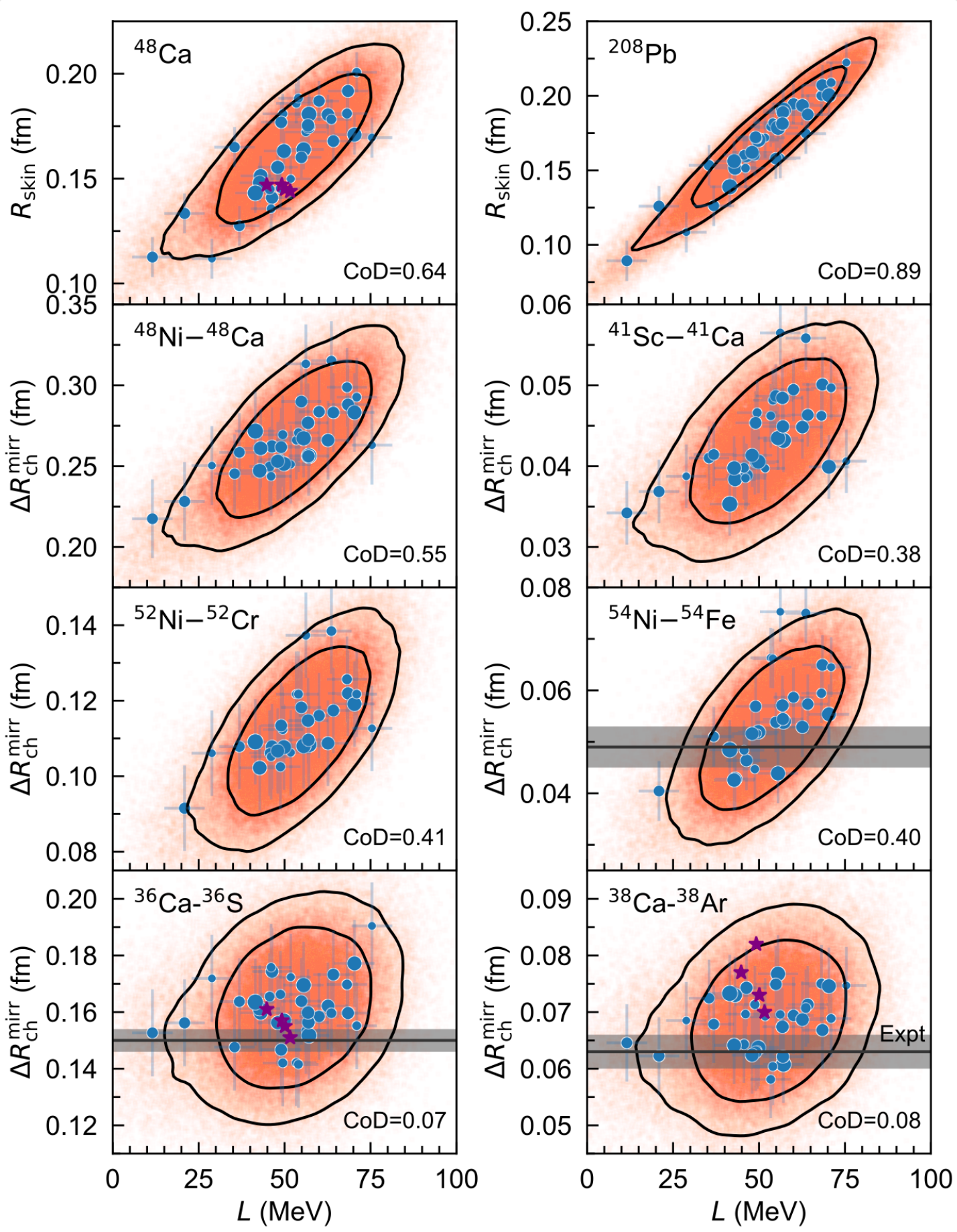}
\caption{\label{Rskin_Rch} Posterior predictive distribution for the slope of symmetry energy at saturation density ($L$), neutron skin thickness ($R_{\rm skin}$), and differences in the charge radii of mirror nuclei ($\Delta R^{\rm mirr}_{\rm ch}$). The sampled bivariate distributions (orange squares) are shown with 68\% and 90\% credible regions (black lines). Blue dots illustrate predictions with the 34 non-implausible chiral interaction samples before error sampling, with sizes proportional to the importance weights for each sample. The theoretical error bars on these blue dots estimate uncertainties from chiral EFT expansion truncations and employed method approximations. The `CoD' refers to the coefficient of determination obtained from these blue dots. The purple stars denote the results based on chiral N$^3$LO NN + N$^2$LO 3N interaction taken from Ref.~\cite{PhysRevResearch.2.022035}. Experimental values for charge radius $R_{\rm ch}$ are taken from Refs~\cite{Rch_2004,Miller2019,PhysRevLett.127.182503}. Note that number of 3 and 6 interaction samples do not yield converged $\Delta R^{\rm mirr}_{\rm ch}$ calculations for A=52 and A=54 pairs, respectively.}
\end{figure*}

Figure~\ref{Rskin_Rch} illustrates  posterior predictive distributions for $R_{\rm skin}$ and $\Delta R ^{\rm mirr}_{\rm ch}$ as a function of $L$. 
In general, starting from the 34 non-implausible chiral interaction samples, the ab initio calculations for $\Delta R ^{\rm mirr}_{\rm ch}$ by employing sampling/importance resampling provide a good description of the existing data, as shown in Table~\ref{table_error}. Fig.~\ref{Rskin_Rch} also displays the MR-IMSRG results for $\Delta R^{\rm mirr}_{\rm ch}$ and $R_{\rm skin}$, which are achieved with an EM family of chiral N$^3$LO NN + N$^2$LO 3N forces \cite{PhysRevResearch.2.022035}. These MR-IMSRG points reside within the 68\% credible region created from the 34 samples, providing a consistent ab initio prediction.
Regarding the neutron skin thicknesses in $^{208}$Pb, a detailed comparison with different measurements is given in Ref.~\cite{Hu22208Pb}. It shows that the ab initio results obtained from the 34 chiral interaction samples are consistent with extractions using electromagnetic and hadronic probes, while being in mild tension (approximately 1.5$\sigma$) with recent PREX results. Furthermore, the results of $^{48}$Ca $R_{\rm skin}$ is in agreement with the recent CREX extraction \cite{PhysRevLett.129.042501}. We do not display experimental data of $R_{\rm skin}$ in Fig.~\ref{Rskin_Rch} and instead focus on the $\Delta R^{\rm mirr}_{\rm ch}$.

As mentioned above, we perform the VS-IMSRG variant developed in Ref.~\cite{PhysRevC.102.034320} to decouple the $^{28}$Si-core $s_{1/2}d_{3/2}f_{7/2}p_{3/2}$ multishell valence space Hamiltonian for the $A$=38 and $A$=41 pairs. This results in $\Delta R ^{\rm mirr}_{\rm ch}$ values of 0.068$\pm$0.007 and 0.043$\pm$0.006 fm, respectively. However, when employing the $sd-$shell and $fp-$shell valence space for $A$=38 and $A$=41 pairs, the $\Delta R ^{\rm mirr}_{\rm ch}$ is 0.089$\pm$0.010 and 0.042$\pm$0.004 fm, respectively. It is clear that the $sd$-shell VS-IMSRG overestimates the data (0.063$\pm$0.003 fm) of $A=38$ $\Delta R ^{\rm mirr}_{\rm ch}$ \cite{Rch_2004,Miller2019}, and it also gives a larger result than the multishell calculation.
As discussed in Ref.~\cite{PhysRevC.102.034320}, excitations from $sd$ to $pf$ are non-negligible for the lighter calcium isotopes. Using the same EM family of chiral interactions, we check that the $^{28}$Si-core $s_{1/2}d_{3/2}f_{7/2}p_{3/2}$ multishell valence space results for the $A$=38 pairs reproduce the MR-IMSRG calculations well, while the $sd$-shell VS-IMSRG overestimates the $\Delta R ^{\rm mirr}_{\rm ch}$. We conclude that the excitations from $sd$ to $pf$ are crucial for describing the $\Delta R ^{\rm mirr}_{\rm ch}$ of $A$=38 pairs.

Figure~\ref{Rskin_Rch} shows that the 68\% credible region of sampled bivariate distribution and CoD for $R_{\rm skin}$ and $L$ are narrower and larger than those for $\Delta R^{\rm mirr}_{\rm ch}$ and $L$, respectively. This indicates that the $R_{\rm skin}-L$ correlation is stronger than the $\Delta R^{\rm mirr}_{\rm ch}-L$ correlation. Notably, $^{208}$Pb $R_{\rm skin}$ exhibits the strongest correlation with $L$ in these calculations. The $fp$-shell mirror pairs with $A=41, 48, 52$ and $54$ yield CoD values of $\Delta R^{\rm mirr}_{\rm ch}-L$ between 0.38 and 0.55, suggesting appreciable correlations that are comparable to the $^{48}$Ca $R_{\rm skin}-L$ correlation. The latest mean-field $\Delta R^{\rm mirr}_{\rm ch}-L$ calculation \cite{PhysRevC.105.L021301} reports CoDs of 0.69 and 0.44 for $A=54$ mirror pair with two different treatments of pairing correlations, which are significantly reduced compared to previous mean-field calculations due to the inclusion of pair correlations and more isovector parameters than relativistic energy density functionals. The ab initio result for $A=54$ pair gives a similar level of CoD with the latest mean-field calculation \cite{PhysRevC.105.L021301}. However, in our results, the CoD is almost zero for $sd$-shell mirror pairs with $A=36$ and $A=38$, indicating that there is no $\Delta R^{\rm mirr}_{\rm ch}-L$ correlation. This finding differs from previous studies ~\cite{PhysRevLett.119.122502,PhysRevResearch.2.022035,PhysRevC.105.L021301}.

Based on results in $sd$-shell and $fp$-shell pairs, we conclude that the correlation between $\Delta R ^{\rm mirr}_{\rm ch}$ and $L$ is significantly influenced by quantum many-body effects. This implies that the formation of neutron skin is also significantly affected by quantum many-body effects, and not all $R_{\rm skin}$ in neutron-rich nuclei exhibit a strong $R_{\rm skin}-L$ correlation. 
We confirm this from CoD of $R_{\rm skin}-L$, which is 0.18, 0.13 for the $sd$-shell $^{36}$S, $^{38}$Ar, respectively, and 0.22, 0.46, 0.31 for the $fp$-shell $^{41}$Ca, $^{52}$Cr and $^{54}$Fe, respectively.
Furthermore, we perform calculations by subtracting the Coulomb interaction and find that it does not change the neutron skin thickness. 

As described in the Methods section regarding the sampling/importance resampling procedure, we assume that the correlation coefficient $R$ between the errors, $\epsilon_{\chi\text{EFT}}$ and $\epsilon_{\text{method}}$, is identical to the correlation coefficient of the observables computed from the 34 non-implausible samples. To validate this assumption, we apply the EFT convergence model \cite{PhysRevC.92.024005,PhysRevC.100.044001,PhysRevC.105.014005} to correlation coefficient $R_{ij}$ for $\epsilon_{\chi\text{EFT}}$, quantifying covariance matrix $\Sigma_{ij}=\bar{c}^2 \cdot y_{\rm ref,i} \cdot y_{\rm ref,j} \cdot \frac{Q^{k+1}_iQ^{k+1}_j}{1-Q^2} \cdot R_{ij}$ \cite{PhysRevC.100.044001,PhysRevC.105.014005}. Here, $\bar{c}$ and the correlation coefficient $R_{ij}$ are learned from the distribution of $c_0$, $c_2$ and $c_3$ in Eq.~(\ref{yk}) for results using LO, NLO and NNLO interactions, respectively. For $\epsilon_{\text{method}}$, we estimate a correlation coefficient of $R_{ij}=0.79$ based on the differences between IMSRG and coupled cluster results versus differences between many-body perturbation theory and coupled cluster results for $R_{\rm skin}$ of $^{208}$Pb, as reported in the extended data Fig. 4 of Ref.~\cite{Hu22208Pb}. Subsequently, we obtain CoD values of 0.54, 0.84, 0.39, 0.36, 0.31, 0.35, 0.17 and 0.12, which compare to those from Fig.~\ref{Rskin_Rch}: CoD = 0.64, 0.89, 0.55, 0.38, 0.41, 0.40, 0.07 and 0.08 for $^{48}$Ca, $^{208}$Pb $R_{\rm skin}-L$, and A=48, A=41, A=52, A=54, A=36, A=38 $\Delta R ^{\rm mirr}_{\rm ch}-L$, respectively. This comparison suggests that the correlation between the errors does not significantly impact the CoD.

\begin{figure}
\centering
\setlength{\abovecaptionskip}{0pt}
\setlength{\belowcaptionskip}{0pt}
\includegraphics[scale=0.21]{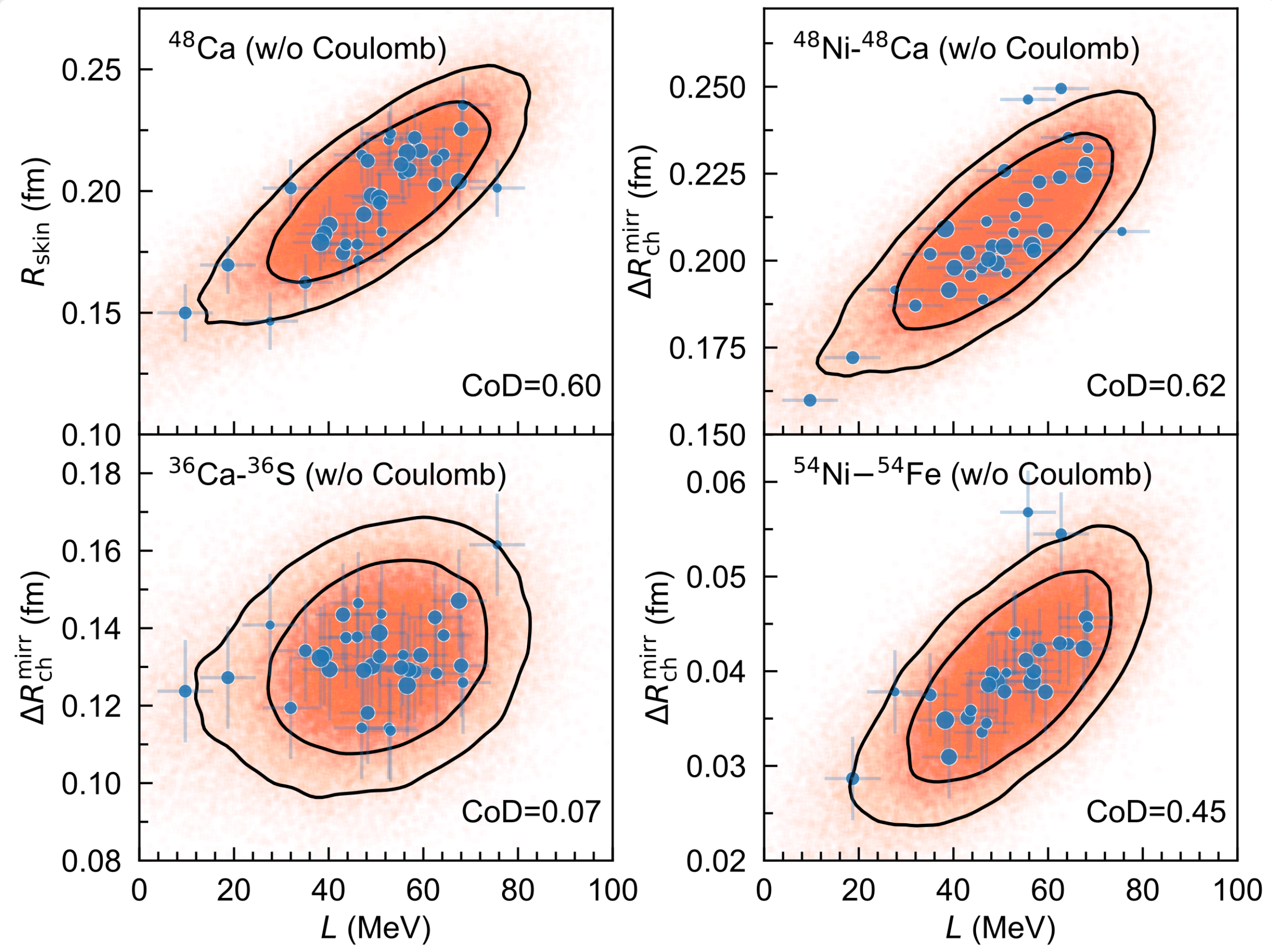}
\caption{\label{Rskin_Rch_woCoul} Similar to Fig.~\ref{Rskin_Rch}, but for calculations of $R_{\rm skin}$ in $^{48}$Ca and $R_{\rm ch}^{\rm mirr}$ in $^{36}$Ca-$^{36}$S, $^{48}$Ni-$^{48}$Ca and $^{54}$Ni-$^{54}$Fe pairs without Coulomb potential.}
\end{figure}

Recently, mean-field studies \cite{PhysRevC.106.L061306,PhysRevC.107.064302} have shown that isospin-symmetry breaking (ISB) has a non-negligible influence on both $\Delta R ^{\rm mirr}_{\rm ch}$ and $R_{\rm skin}$. We refer the reader to the paper~\cite{PhysRevC.107.064302} for an overview of the ISB effects on various nuclear properties. To investigate the ISB effects arising from chiral interactions \cite{Machleidt20111}, we calculate the $\Delta R ^{\rm mirr}_{\rm ch}$ and $R_{\rm skin}$ with and without Coulomb interaction. For $\Delta R^{\rm mirr}_{\rm ch}$, Fig.~\ref{Rskin_Rch_woCoul} presents results without Coulomb potential for the $^{36}$Ca$-^{36}$S, $^{48}$Ni$-^{48}$Ca and $^{54}$Ni$-^{54}$Fe pairs. Comparing these results with the corresponding ones in Fig.~\ref{Rskin_Rch}, which include the Coulomb potential, we find that the Coulomb interaction does not change the $\Delta R^{\rm mirr}_{\rm ch}-L$ correlation, although it increases the $\Delta R ^{\rm mirr}_{\rm ch}$ through repulsive effects. Fig.~\ref{Rskin_Rch_woCoul} also displays the $^{48}$Ca $R_{\rm skin}$ result without the Coulomb potential. Coulomb interaction does not contribute to the $R_{\rm skin}-L$ correlations, although it decreases the $R_{\rm skin}$. By comparing the $^{48}$Ca $R_{\rm skin}$ and $^{48}$Ni$-^{48}$Ca $R_{\rm ch}^{\rm mirr}$, we see that the results are very similar, with the small difference mainly arising from ISB terms in chiral interactions. In the calculation for unbound $^{48}$Ni, we have not included continuum coupling in the present framework, but we plan to include it in future developments. Studies in Ref.~\cite{PhysRevC.106.L061306} claim that the continuum effects are small due to the Coulomb plus centrifugal barrier felt by the protons is large enough. It is worth noting that Refs.~\cite{PhysRevC.104.014324,PhysRevC.107.014302} have explored the ISB in mirror energy difference, nuclear mass, and spectra using chiral interactions within the VS-IMSRG framework.

\section{Summary}
In this work, using recently developed 34 chiral NN+3N interaction samples identified by the history matching approach, we probed the possible correlations among the differences in the charge radii of mirror nuclei ($\Delta R^{\rm mirr}_{\rm ch}$), neutron skin of neutron-rich ($R_{\rm skin}$), and the slope of the symmetry energy ($L$) by performing valence-space in-medium similarity renormalization group approach. We also analyzed the isospin-symmetry-breaking effect by subtracting Coulomb potential.

This letter has shown that the $\Delta R^{\rm mirr}_{\rm ch}$ does not provide as stringent constraints on $L$ as $R_{\rm skin}$ of $^{208}$Pb. However, in the medium-mass region,  $\Delta R^{\rm mirr}_{\rm ch}$ could exhibit a comparable correlation with $L$ as the $R_{\rm skin}$ in $^{48}$Ca. Contrary to the previous conclusions, our ab initio results reveal that $sd$-shell mirror pairs with $A=36$ and $A=38$ do not exhibit any correlation with $L$. We find that the $\Delta R^{\rm mirr}_{\rm ch}-L$ correlation is significantly influenced by the quantum many-body effects, which also play a crucial role in the formation of neutron skin. We conclude that the $\Delta R^{\rm mirr}_{\rm ch}$ could provide an additional perspective that is crucial for a complete test of nuclear forces and many-body methods. 
Work is currently in progress to perform ab initio examinations of potential correlations between other quantities by accounting for quantified uncertainties from both chiral EFT truncation errors and many-body method approximation uncertainties.

\section*{Acknowledgements}
We are grateful to J.D. Holt for valuable discussions and support, and to T. Papenbrock for the useful discussion and reading of this article. We thank A. Belley, C. Forss{\'e}n, G. Hagen, W. G. Jiang and T. Miyagi for their helpful discussions.
TRIUMF receives funding via a contribution through the National Research Council of Canada.
TRIUMF receives funding via a contribution through the National Research Council of Canada.
This work was further supported by the Arthur B. McDonald Canadian Astroparticle Physics Research Institute and the U.S. Department of Energy, Office of Science, under SciDAC-5 (NUCLEI collaboration). 
The IM-SRG code used is Ragnar$\_$IMSRG \cite{ragnar}.
Computations were performed with an allocation of computing resources on Cedar at WestGrid and Compute Canada, and on the Oak Cluster at TRIUMF managed by the University of British Columbia Department of Advanced Research Computing (ARC). This research also used resources of the Oak Ridge Leadership Computing Facility located at Oak Ridge National Laboratory, which is supported by the Office of Science of the U.S. Department of Energy under contract No. DE-AC05-00OR22725.



  \bibliographystyle{elsarticle-num_noURL}
  \bibliography{references}





\end{document}